\documentclass[
superscriptaddress, twocolumn,
 amsmath,amssymb,
 aps,
 pre,
floatfix,
]{revtex4-1}

\usepackage{graphicx}
\usepackage{bm}
\usepackage{amsmath}
\usepackage{mathrsfs}
\usepackage{appendix}
\usepackage{bbold}

\begin{document}


\title{Geometry-based circulation of local thermal current in quantum harmonic and Bose Hubbard systems}

\author{Palak Dugar}
\affiliation{Department of physics, University of California, Merced, CA 95343, USA.}
\author{Chih-Chun Chien}%
\email{cchien5@ucmerced.edu}
\affiliation{Department of physics, University of California, Merced, CA 95343, USA.}


\date{\today}

\begin{abstract}
A geometry-based mechanism for generating steady-state internal circulation of local thermal currents is demonstrated by harmonically coupled quantum oscillators formulated by the Redfield quantum master equation (RQME) and the Bose Hubbard model (BHM) of phonons formulated by the Lindblad quantum master equation (LQME) using the simple multi-path geometry of a triangle. Driven by two reservoirs at different temperatures, both systems can exhibit an atypical local thermal current flowing against the total current. However, the total thermal current behaves normally. While the RQME of harmonically coupled quantum oscillators allows an analytical solution, the LQME of the interacting BHM can be solved numerically. The emergence of the geometry-based circulation in both systems demonstrates the ubiquity and robustness of the mechanism. In the high-temperature limit, the results agree with the classical results, confirming the generality of the geometric-based circulation across the quantum and classical boundary.
Possible experimental implications and applications are briefly discussed.
\end{abstract}

\maketitle

\section{\label{sec:level1}Introduction} 

Geometry plays a crucial role in transport phenomena \cite{wang2021geometric}. For example, the ring geometry provides a natural shape for persistent current induced by magnetic flux \cite{maiti2014quantum,Jayich09,castellanos2013measurement}.  Interestingly, circulation of electrons in the form of
current vortices \cite{stegmann2020current, arnold2013magnetic, barquera2019vorticity, gomes2021current} or chiral current \cite{downing2020chiral} may emerge without an external magnetic field. A metallic ring embedded with two quantum dots and connected to external electrodes shows circulating currents \cite{anda2012circulating}. Ref. \cite{cho2005thermal} shows the
quantum interference of tunneling electrons in two quantum dots individually coupled to two reservoirs at different temperatures, resulting in a circulating electric current with magnetic polarization. 
Moreover, geometry-based circulations of electrons~\cite{lai2018tunable} and photons~\cite{dugar2020geometry} have been predicted in quantum dot or photonic structures, showing that geometric effects transcends spin-statistics. Topological properties, such as defects in 1D Bose fields, may also be investigated in a ring geometry \cite{roscilde2016quantum}.

Meanwhile, thermal transport in classical and quantum systems has been intensively studied \cite{kato2016quantum, dhar2008heat,ren2010emergence, zhu2016persistent,iubini2018heat}. Interesting phenomena, including heat rectification in spin systems \cite{balachandran2019heat}, heat flux from the nontrivial Berry-phase in an anharmonic molecular junction subjected to cyclic modulations, \cite{ren2010berry}, a local thermal current from cold to hot in a multi-path electronic system \cite{cho2005thermal}, have been studied. Ref. \cite{dugar2019geometry} shows that classical harmonic systems in a multi-path geometry support local atypical thermal current flowing from cold to hot in the steady state. Here we investigate the quantum version of Ref. \cite{dugar2019geometry} and its generalizations to establish the ubiquity of the geometry-based circulation in thermal transport, showing the geometric mechanism transcends the quantum and classical boundary. One important advantage of the geometric mechanism is the lack of a direct means for stirring thermal currents, in contrast to the electrons or photons that may be manipulated by a magnetic field~\cite{griffiths1999introduction} or artificial gauge field \cite{hey2018advances, fang2012photonic}.

The two examples in this work for demonstrating the geometry-based circulation in quantum thermal transport are formulated by the quantum master equation (QME) that describes the time evolution of the reduced density matrix of a system coupled to external reservoirs. Depending on the approximations in the derivations that will be presented later, we analyze the Redfield quantum master equation (RQME) \cite{purkayastha2017quantum,lidar2019lecture} of harmonically coupled quantum oscillators and the Lindblad quantum master equation (LQME) \cite{lidar2019lecture, breuer2002theory} of the Bose Hubbard model (BHM) of phonons. Here the phonons refer to the energy quanta of the underlying oscillators, not the phonons in crystals.
Challenges arise when solving the QMEs due to the rapid growth of the Fock space with the particle number. Exact numerical simulations are not practical as the system size increases. For harmonic oscillators or noninteracting BHM, we implementing the third quantization method of bosons~\cite{prosen2010quantization,vzunkovivc2012heat}. This method allows an analytical evaluation of the non-equilibrium steady-state correlations, from which the local and total thermal currents can be obtained.
To explore interaction effects, we implement numerical calculations of the interacting BHM with a truncated Fock space, similar to those of photonic transport~\cite{dugar2020geometry}. We mention that quantum thermal transport may also be studied by using non-equilibrium Green's function \cite{wang2008quantum}, quantum Langevin equation \cite{attal2007langevin, dhar2006heat}, quantum stochastic Schr\"{o}dinger equation \cite{zoller1997quantum}, quantum master equations \cite{asadian2013heat}, and many others~\cite{Landi21}. 

From the different QMEs of different models, we will establish the geometry-based circulation of steady-state thermal currents in the quantum regime. Similar to the classical harmonic systems~\cite{dugar2019geometry}, the circulation of the quantum thermal current is a consequence of the competition among the local thermal currents carried along different paths. For the classical harmonic systems, the mechanical vibrational modes carry the energy while for the quantum system, the wavefunctions of the phonons transport the energy. Moreover, the geometry-based circulation in quantum systems is shown to be robust against interactions, just like the circulation in classical systems is robust against nonlinear potentials~\cite{dugar2019geometry}.

The rest of the paper is organized as follows. In Section \ref{sec:level2}, we describe the two systems with their Hamiltonians, the quantum master equations for their time evolution, the definitions of the local and total thermal currents, and the methods for obtaining the steady-state results. In Section \ref{sec:level3}, we present the local and total thermal currents of both systems in the steady state, establishing the  geometry-based circulation of thermal currents in quantum transport. The patterns and the phase diagrams will be presented. Section \ref{sec:exp} discusses experimental implications and possible applications. We conclude our study in Section \ref{sec:level4}. Some details of the RQME and connections between the LQME and the RQME are summarized in the Appendix.   


\section{\label{sec:level2}Models and Methods}
\subsection{Quantum harmonic oscillators with RQME}
\begin{figure}[t]
	\centering
	\includegraphics[width=0.49\textwidth]{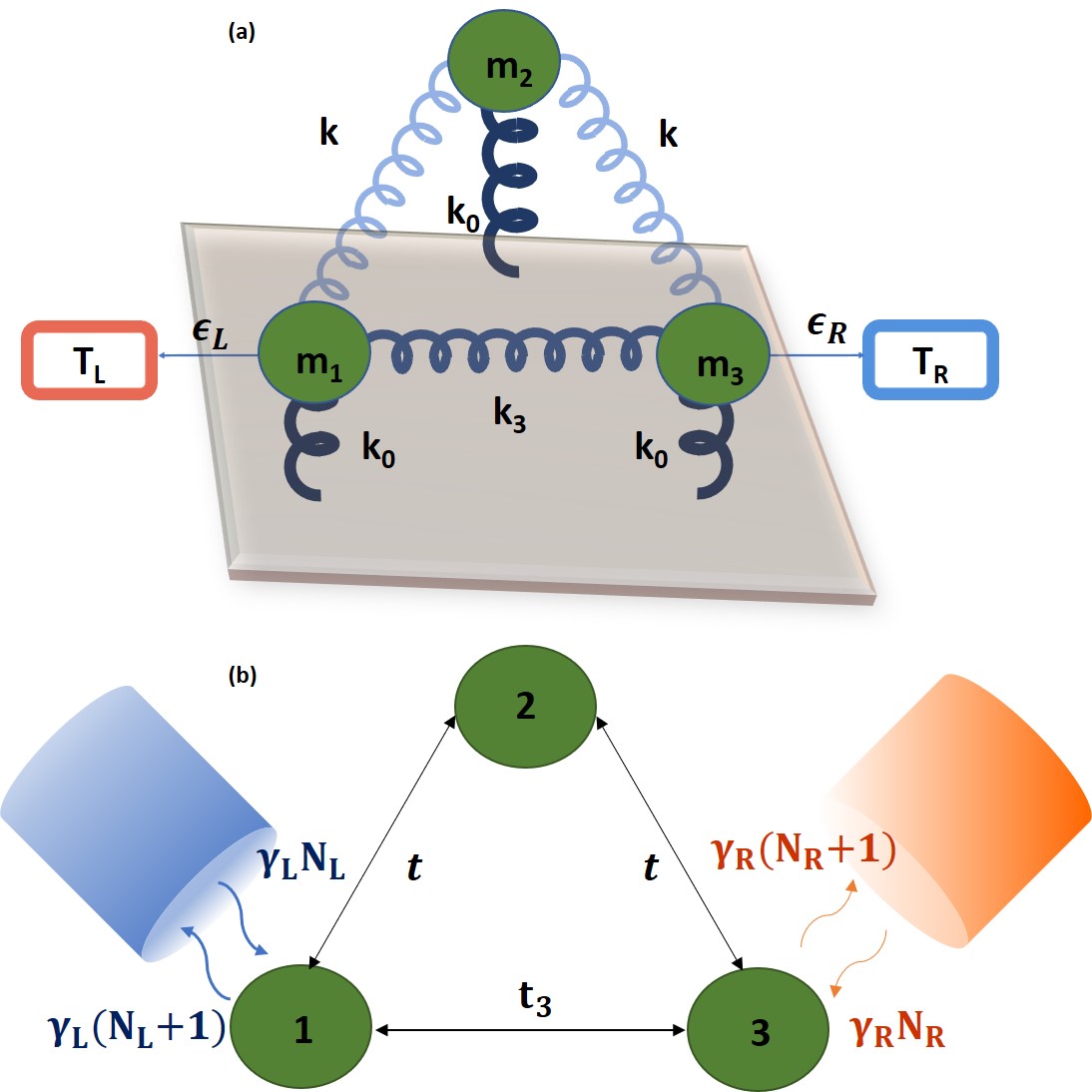}
	\caption{\label{fig:schematic_red_lin} Schematic illustrations of the systems for studying geometry-based circulation in quantum thermal transport. (a) Harmonically coupled quantum oscillators driven by two reservoirs with temperatures $T_{L}$ and $T_{R}$, respectively. The harmonic coupling constants between $m_{2}$-$m_{1}$ and $m_{2}$-$m_{3}$ are set to $k$ while that between $m_{1}$-$m_{3}$ is set to $k_3$. All masses are coupled to the substrate with harmonic coupling constant $k_0$. (b) Bose-Hubbard model of phonons with tunneling coefficients $t$ along the 1-2 and 2-3 links and $t_3$ along the 1-3 link. The system is connected via the system-reservoir couplings $\gamma_L$ and $\gamma_R$ to two reservoirs with temperatures $T_L$ and $T_R$ and average numbers $N_L$ and $N_R$, respectively.}
\end{figure}

To investigate local thermal transport in a multi-path geometry, we consider a simplified system of three quantum oscillators with equal mass $m$ harmonically coupled to each other and to a substrate, as shown in Fig.~\ref{fig:schematic_red_lin} (a).  Following Ref. \cite{vzunkovivc2012heat}, the Hamiltonian in the mass weighted coordinates can be written as
\begin{equation}\label{eq:QOham}
\begin{split}
\mathscr{H} =& \frac{1}{2}\sum_{j=1}^3 (p_j^2 + \omega_{0}^2q_{j}^2) + \frac{k_3}{2m}(q_1-q_3)^2 \\ & +\frac{k}{2m}\sum_{j=1}^2 (q_j-q_{j+1})^2 \nonumber \\
=& \frac{1}{2}(p.p + q.\mathbf{Q}q).
\end{split}
\end{equation}
Throughout the paper, $\hbar=k_B=1$.  Here $q_j$ and $p_j$, $j=1,2,3$, denote the displacement and momentum operators of the $j$-th oscillator, $\omega_0$ is the frequency from the uniform onsite potential associated with the substrate. The harmonic coupling constants are $k$ between the $m_1-m_2$ link and the $m_2-m_3$ link and $k_3$ between the $m_1-m_3$ link.
The $p$ and $q$ are column vectors and
\begin{equation}\label{eq:QOham_in_Q}
\mathbf{Q} = \omega_{0}^2\mathbb{1}_3 +\omega_{c}^2
\begin{pmatrix}
1+\frac{k_3}{k} & -1 & -\frac{k_3}{k} \\
-1 & 2 & -1 \\
-\frac{k_3}{k} & -1 & 1+\frac{k_3}{k}
\end{pmatrix},
\end{equation}
with $\omega_c=\sqrt{k/m}$. 
In principle, the masses, onsite frequencies, and harmonic coupling constants are all tunable, giving rise to rich physics.

To study thermal transport, the quantum system is connected to two thermal reservoirs maintained at temperatures $T_{L,R}$, respectively. Without loss of generality, we assume $T_L>T_R$.
The time evolution of the reduced density matrix of the system under the influence of the reservoirs may be described by the RQME~\cite{vzunkovivc2012heat}:
\begin{equation}\label{eq:RQME}
\frac{d\rho(\mathcal{T})}{d\mathcal{T}}=\iota[\rho(\mathcal{T}), \mathscr{H}] +\mathscr{D}\rho(\mathcal{T}).
\end{equation}
Here $\mathscr{D}$ is the Redfield dissipator given by
\begin{equation}\label{eq:redfielddiss}
\mathscr{D}_{L,R}\rho(\mathcal{T}) =\int_0^\infty d\tau \Gamma_{L,R}(\tau)[X_{L,R}(-\tau)\rho(\mathcal{T}),X_{L,R}] + h.c.
\end{equation}
$[A,B]$ represents the commutator of operators $A$ and $B$. 
For the quantum oscillator system, the coupling operators are \cite{vzunkovivc2012heat}  $X_{L,R}= \sqrt{\epsilon_{L,R}}q_{1,3}$,  respectively. Here $\epsilon_L$ and $\epsilon_R$ are the system-reservoir coupling constants. Ref. \cite{vzunkovivc2012heat} presents a general form of the spectral function of the thermal reservoir, we focus on the reservoirs with an ohmic spectral function, which is the Fourier transform of $\Gamma_{L,R}(\tau)$, defined as
\begin{equation}\label{eq:specfunkinfreqspace}
\Gamma_{L,R}(\omega)=\frac{sign(\omega)|\omega|}{\exp{(\omega/T_{L,R})}-1}.
\end{equation}

The Redfield form of the master equation is derived using the Born and Markov approximations. Under the Born approximation \cite{lidar2019lecture, breuer2002theory} it is assumed that the reservoirs are large compared to the system such that the reservoir is not affected significantly because of the reservoir-system interaction and the interaction between the system and the reservoirs is weak. With the Born approximation, the reduced density matrix of the system and the density matrix of the reservoirs is written as a product state. Under the Markov approximation, it is assumed that the timescale of the system dynamics is larger than the reservoir correlation time, allowing the master equation to be written as a time local equation. 

In the long-time limit, the system is expected to reach a steady state. The third quantization formalism provides a method for solving the RQME with a quadratic Hamiltonian like Eq.~\eqref{eq:QOham} subject to linear system-reservoir coupling operators ~\cite{vzunkovivc2012heat}. For the harmonically coupled quantum oscillators, we extract the non-equilibrium steady state (NESS) momentum-coordinate correlations from the RQME with the details summarized in Appendix \ref{appendix}.
The local thermal current from the $i$-th oscillator to its adjacent $j$-th oscillator is derived from the continuity equation \cite{prosen2010exact}, given by
\begin{equation}\label{eq:J_ij_SHO}
\langle J_{ij} \rangle =\frac{K}{m}tr{(p_jq_i\rho)},
\end{equation}
where $K$ denotes the harmonic coupling constant between the two sites. For a harmonic system in the NESS, $\langle J_{12} \rangle = \langle J_{23} \rangle$ with $\rho=\rho_{NESS}$, so there is no energy accumulation in oscillator $2$ that is not coupled to a reservoir. The total steady-state thermal current through the system is given by
\begin{equation}
\langle J_{T} \rangle = \langle J_{13} \rangle + \langle J_{12} \rangle.
\end{equation}

\subsection{Bose Hubbard model with LQME}
A quantum oscillator can be expressed in terms of the creation and annihilation operators of energy quanta \cite{sakurai2014modern}. It can be shown that under the rotating wave approximation neglecting number non-conserving terms, a system of harmonically coupled quantum oscillators may be approximated by the non-interacting BHM of phonons~\cite{asadian2013heat, nicacio2015thermal}. The approximation may be justified when the onsite frequency dominates the hopping coefficients. From here on, we investigate quantum thermal transport through a BHM in a multi-path geometry illustrated in Fig.~\ref{fig:schematic_red_lin} (b). Our goal is not to establish a rigorous connection to the coupled quantum oscillators but instead to demonstrate the robustness of the geometry-based circulation of local thermal current in open quantum systems.

The three-site BHM has the following Hamiltonian:
\begin{equation}\label{eq:BHHam}
    \begin{split}
    \mathscr{H}_{BH} & = \sum_{j=1}^3 \Omega_{0}c_{j}^{\dagger}c_{j} 
    -t(c_{1}^{\dagger}c_{2} + c_{2}^{\dagger}c_{1} + c_{2}^{\dagger}c_{3} + c_{3}^{\dagger}c_{2}) \\
    &-t_{3}(c_{1}^{\dagger}c_{3} + c_{3}^{\dagger}c_{1})
    +\frac{U}{2}\sum_{j=1}^3 n_{j}(n_{j}-1).  
    \end{split}
\end{equation}
Here a uniform onsite potential $\Omega_{0}$, possibly from the system-substrate coupling, has been included, $c_{j}^{\dagger}$ and $c_{j}$ are the creation and annihilation operators at the $j^{th}$ site. $t$ is the tunneling coefficients between site $1$-site $2$ and site $2$-site $3$, and $t_{3}$ is the tunneling coefficient between site $1$-site $3$. We will focus on the regime  with $\Omega_0>>t,t_3$. $U$ is the onsite coupling constant and $n_j$ is the number density operator on site $j$. We will begin with the non-interacting case with $U=0$ and consider the interacting case afterwards. 
The LQME~\cite{haroche2006exploring} will be implemented to study the local thermal currents in the BHM driven by two thermal reservoirs. One may derive the LQME from the Kraus operator formalism \cite{haroche2006exploring} and utilize it as a phenomenological equation \cite{jeske2015bloch, gao1997dissipative,lindblad1976brownian, breuer2002theory}.  
On the other hand, one may derive a limited class of the LQMEs \cite{breuer2002theory, lidar2019lecture, wichterich2007modeling} by imposing either the secular approximation or the weak internal-coupling approximation on the RQME. The derivation and approximation of the LQME limit the parameter space where it can be used appropriately. Here we use the LQME as a phenomenological equation that allows us to explore a broad region in the parameter space. More discussions on the limitations of the LQME have been summarized in Ref.~\cite{Tupkary21}. For the sake of completeness, we briefly mention the LQMEs obtained from the RQME in Appendix~\ref{app:QME}.

The LQME we work with has the form:
\begin{eqnarray}\label{eq:LQME}
\partial{\rho(\mathcal{T})}/\partial{\mathcal{T}}&=& \iota[\rho,\mathscr{H}_{BH}]+ \gamma_{L}N_{L}(c_{1}^{\dagger}\rho c_{1} -\frac{1}{2}\{c_{1}c_{1}^{\dagger},\rho\})  \nonumber \\ 
& &+ \gamma_{L}(N_{L}+1)(c_{1}\rho c_{1}^{\dagger} -\frac{1}{2}\{c_{1}^{\dagger}c_{1},\rho\})  \nonumber \\ 
& &+ \gamma_{R}N_{R}(c_{3}^{\dagger}\rho c_{3} -\frac{1}{2}\{c_{3}c_{3}^{\dagger},\rho\})  \nonumber \\ 
& &+ \gamma_{R}(N_{R}+1)(c_{3}\rho c_{3}^{\dagger} -\frac{1}{2}\{c_{3}^{\dagger}c_{3},\rho\}).
\end{eqnarray}
Here $\{A,B\}$ represents the anti-commutator of operators $A$ and $B$. $\gamma_{L}$ and $\gamma_{R}$ are the system-reservoir couplings for the left and right reservoirs that are assumed to maintain fixed phonon numbers $N_L$ and $N_R$ with $N_{L,R}=1/[\exp(\Omega_0/T_{L,R})-1]$, respectively. The reservoirs emit phonons at the rate  $\gamma_{j}N_{j}$ into the system while they absorb at the rate $\gamma_{j}(N_{j}+1)$ with $j=L, R$, as shown in Eq.~\eqref{eq:LQME}. These exchange rates of phonons follow the assumption of Bose statistics and lead  the system to thermal equilibrium if only connected to a single reservoir. 

There have been concerns about the thermodynamic consistency of the local (position basis) LQME~\cite{purkayastha2016out, cattaneo2019local, stockburger2017thermodynamic, levy2014local}. However, those issues have been addressed by choosing the correct thermodynamic definitions of the currents related to the work and heat \cite{de2018reconciliation,hewgill2021quantum}, respectively. We will focus on the thermal current associated with heat throughout the paper. Explicitly, there are multiple ways of defining the thermal current in quantum systems~\cite{asadian2013heat,hewgill2021quantum, prosen2010exact}. Ref. \cite{asadian2013heat} shows two different formulae of the thermal current of a linear chain. In the first expression, the QME in the steady state enforces $tr(\mathscr{H}\frac{\partial \rho}{\partial \mathcal{T}})=0$. The commutator in Eq. \eqref{eq:LQME} does not contribute, and the contributions from the two reservoirs sum to zero in the steady state, making them equal and opposite to each other. Without loss of generality, either the contribution from the left or right reservoir may be picked as the thermal current. However, this expression of the thermal current comprises of not only the heat but also the work exchanged at the system-reservoir interface. The subtlety is explained in Ref. \cite{hewgill2021quantum}, which shows that only the diagonal terms of the Hamiltonian contribute to the heat as they are the ones that enter the entropy production. Instead, the non-diagonal terms of the Hamiltonian contribute to the work done on the system.

To focus on the heat transferred through the system, we choose the second expression~\cite{asadian2013heat, prosen2010exact} and derive a formula of the thermal current associated with heat in thermodynamics. For a linear chain described by the BHM, the local thermal current operator associated with heat through a link between the $i^{th}$ and $(i+1)^{th}$ sites can be evaluated by the Heisenberg equation of motion. Explicitly, one defines the Hamiltonian of the partial chain up to the link as $H_L$. Then, $J_{i,i+1}=dH_L/dt=i[H_L,H]$, where $H_L$ contains the Hamiltonian from the left end to the left site of the link. By generalizing the definition, the thermal current operator from site $i$ to its adjacent site $j$ is given by
\begin{equation}\label{eq:localcurrentwithU}
\begin{split}
    J_{ij} &=-\iota t_{ij}(\Omega_{0}-U)(c_{i}^{\dagger}c_{j}-c_{j}^{\dagger}c_{i}) \\& 
    -\iota Ut_{ij}(c_{i}^{\dagger}c_{j}c_{j}^{\dagger}c_{j}-c_{j}^{\dagger}c_{j}c_{j}^{\dagger}c_{i}).
\end{split}
\end{equation}
Here $t_{ij}$ takes the value $t$ or $t_3$ for $J_{12}$ or $J_{13}$, respectively. 
The local thermal current is obtained from $\langle J_{ij}\rangle=tr(\rho J_{ij})$. Using the above definition, we make sure that the total thermal current does not violate the second law of thermodynamics, as it always flows from hot to cold. Nevertheless, atypical local thermal currents will be shown to be able to flow against the total thermal current in quantum systems.


For the noninteracting case with $U=0$, we implement the third quantization method for the LQME \cite{prosen2010quantization, pivzorn2013one} to obtain the steady-state currents with a similar procedure described in Ref. \cite{dugar2020geometry}. 
The third quantization method is limited to systems with quadratic Hamiltonians. When $U$ is finite, the nonlinear terms no longer permit us to utilize the third quantization framework. Therefore, we numerically simulate the LQME of the interacting BHM in a truncated basis. We follow Ref. \cite{szabados2012efficient} to construct the truncated basis states for the system. To numerically integrate Eq. \eqref{eq:LQME}, we use the fourth-order Runge-Kutta method \cite{press2007numerical}, which yields the evolution of the reduced density matrix. After checking the system reaches a steady state in the long-time limit, the expectation values of the local current operators are evaluated.
The local thermal currents of the interacting BHM in the NESS should obey $\sum_{i}\langle J_{i2} + J_{2i}\rangle=0$ to ensure no energy accumulation on site $2$ that is not coupled to a reservoir. The dimension of the density matrix of a bosonic system increases rapidly with the number of bosons, so numerical simulations of bosons are usually performed by restricting the maximal number of bosons per site. In our simulations, we have checked the results with the number of phonon per site up to 4 and only see quantitative differences.

\begin{figure}[t]
	\centering
	\includegraphics[width=0.48\textwidth]{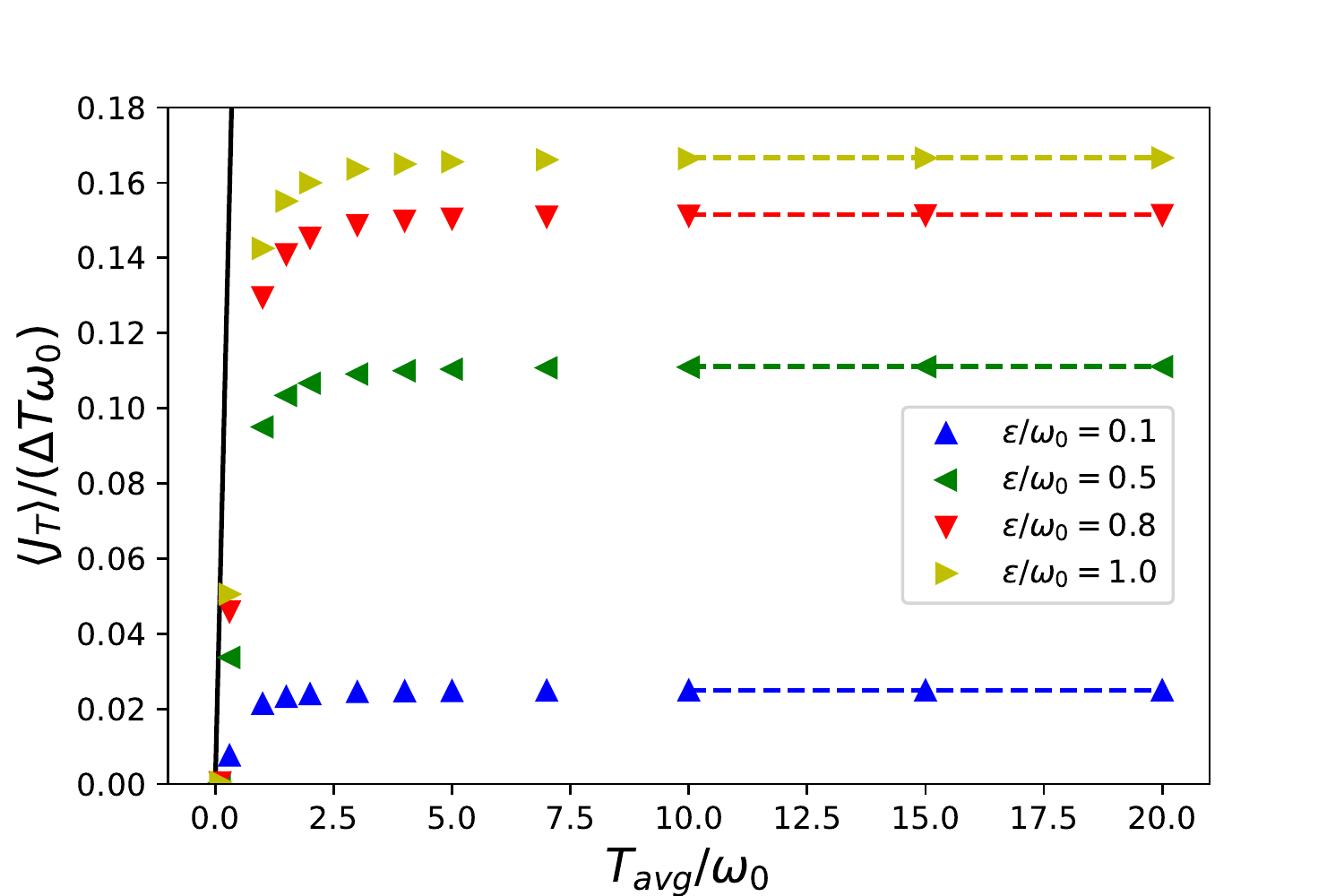}
	\caption{\label{fig:RedconductancevsTavg} Quantum thermal conductance from the total current through the three-site harmonic oscillators with equal mass $m$ and $k=k_3=m\omega_0^2$ as a function of $T_{avg}/\omega_0$ for different values of the system-reservoir coupling and fixed $\Delta T/T_{avg}=0.02$. The black line shows the quantum of thermal conductance and the dashed lines show the corresponding values of the classical thermal conductance.}
\end{figure}

\section{\label{sec:level3}Results and discussions}
\subsection{Quantum oscillators with RQME}\label{res:RQME}
We first present the results of quantum thermal transport through harmonically coupled quantum oscillators described by the Redfield master equation. Before showing the results of the setup shown in Fig.~\ref{fig:schematic_red_lin}, we have verified that our results for a linear chain of quantum oscillators are consistent with the results of Ref.~\cite{vzunkovivc2012heat}. In the following, We choose $T_L-T_R=\Delta T << T_{avg}$, where $T_{avg}=(T_L+T_R)/2$ is the average temperature of the reservoirs and assume symmetric couplings to the reservoirs, $\epsilon_{L,R}=\epsilon$. The parameter space of the system with equal mass $m$ in a triangular geometry consists of the internal parameters $k/k_{3}$ and external parameters $T_L/\omega_0$, $T_R/\omega_0$, and $\epsilon/\omega_0$.

\begin{figure}[t]
	\centering
	\includegraphics[width=0.48\textwidth]{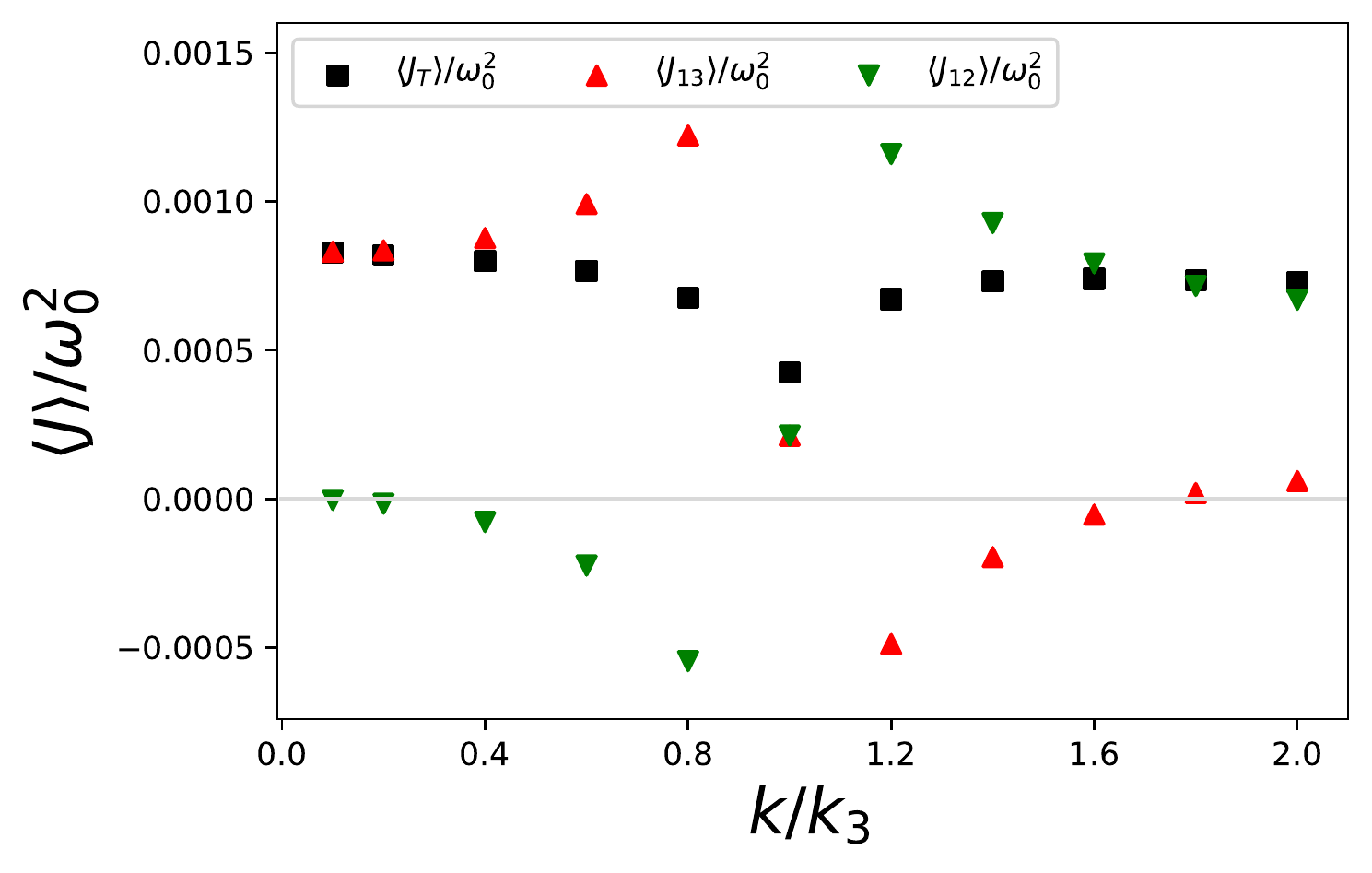}
	\caption{\label{fig:redlocalandtot} Total and local steady-state thermal currents of the three-site quantum oscillators described by the RQME as a function of $k/k_3$ for $\epsilon/\omega_0=0.1$. Here $T_L/\omega_0=1.01$ and $T_R/\omega_0=0.99$. $\omega_0$ is the onsite frequency. The solid grey line marks the zero of the $y$-axis.}
\end{figure}

The third quantization method of the RQME~\cite{vzunkovivc2012heat} gives the steady-state expectation values. The thermal conductance $\langle J_T \rangle/\Delta T$ as a function of $T_{avg}$ for selected values of the system-reservoir coupling $\epsilon$ is shown in Fig. \ref{fig:RedconductancevsTavg} with $k=k_{3}=m\omega_0^2$. 
In the low temperature regime when $\omega_{0}\gg T_{avg}$, the conductance increases monotonically as $T_{avg}$ increases.
The quantum of thermal conductance \cite{dhar2008heat} in the low-temperature limit is given by $\pi {k_{B}}^2 T_{avg}/6 \hbar$ and shown by the
black curve in Fig. \ref{fig:RedconductancevsTavg}. 
As one can see, the quantum of thermal conductance becomes an upper bound for the numerical values as $T_{avg}\rightarrow 0$. When $T_{avg}$ increases, the thermal conductance starts to saturate and becomes constant. At high temperatures, the system shown in Fig.~\ref{fig:schematic_red_lin}(a) approaches a classical mechanical system, and the spectrum of the reservoirs is expected to approach the while noise. Classical thermal transport of Fig.~\ref{fig:schematic_red_lin} (a) with white-noise reservoirs has been studied by us in Ref.~\cite{dugar2019geometry}. The RQME results approach the corresponding classical values in the high-temperature limit, which are shown as the dashed lines in Fig.~\ref{fig:RedconductancevsTavg}. The agreement between the RQME and the classical Langevin results in the high temperature limit $\omega_{0}\ll T_{avg}$ has been shown in a linear chain of quantum harmonic oscillators~\cite{vzunkovivc2012heat}, and here we confirm the agreement in a multi-path geometry.

\begin{figure}[t]
	\centering
	\includegraphics[width=0.48\textwidth]{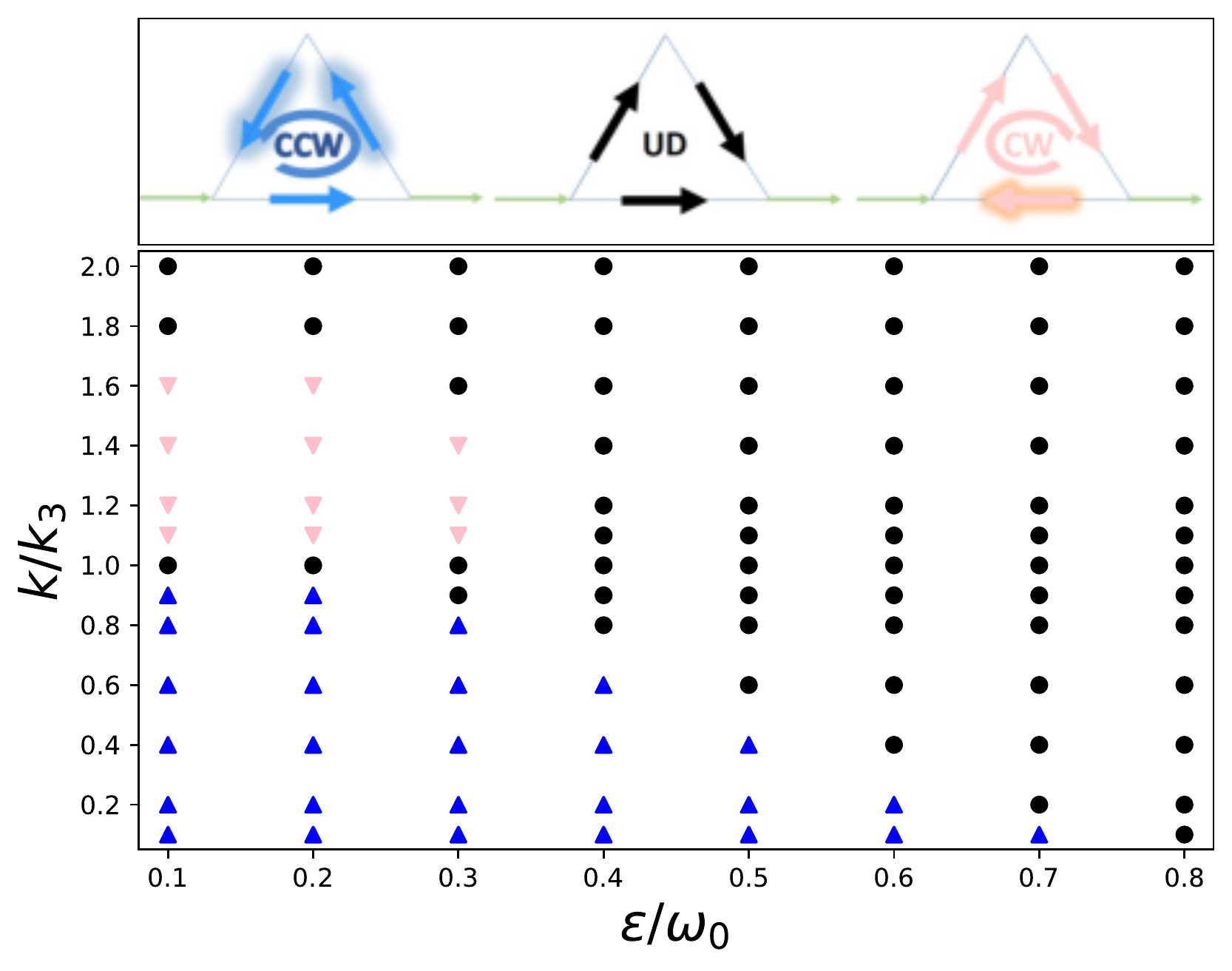}
	\caption{\label{fig:redfield_phase} (Top) Patterns of the local thermal currents. Form left to right: counterclockwise (CCW), uni-directional (UD), and clockwise (CW). (Bottom) Phase diagram showing where each pattern survives as a function of $k/k_3$ and $\epsilon/\omega_0$ for the three-site quantum harmonic oscillators described by the RQME. Here  $T_L/\omega_0=1.01$ and $T_R/\omega_0=0.99$, and the masses are the same with $\sqrt{k_{3}/m}=\omega_0$. The blue triangles, black circles, and pink inverted triangles represent the CCW circulation, unidirectional flow, and CW circulation, respectively. }
\end{figure}

We caution that the RQME is derived as a
second-order perturbation~\cite{vzunkovivc2012heat}, which is reliable when the system-reservoir coupling is weak.
Within the weak coupling regime, we found the thermal conductance increases with $\epsilon/\omega_0$, as shown in Fig. \ref{fig:RedconductancevsTavg}. 
The thermal conductance of a classical harmonic chain has been shown to change non-monotonically with the system-reservoir coupling if the coupling is varied by several orders of magnitude~\cite{velizhanin2015crossover}. However, the weak-coupling assumption of the RQME limits our ability to explore quantum thermal transport in the regime where $\epsilon/\omega_0 > 1$.

The total thermal current does not reveal exciting physics. Nevertheless, we unambiguously demonstrate the existence of atypical local thermal current in the quantum system shown in Fig.~\ref{fig:schematic_red_lin} (a) by presenting the local thermal currents in Fig.~\ref{fig:redlocalandtot} with uniform mass $m$,
$T_L/\omega_0=1.01$, $T_R/\omega_0=0.99$, $k_3=m\omega_0^2$, and $\epsilon/\omega_0=0.1$. As one can see on Fig.~\ref{fig:redlocalandtot}, the local thermal current along the $1-3$ link flows from hot to cold according to the direction of the reservoirs when $k<k_{3}$, but the local thermal current along the $1-2$ link flows from cold to hot as indicated by the negative value. In the steady state, we have verified that $\langle J_{12} \rangle=\langle J_{23} \rangle$. The combination of negative $\langle J_{12} \rangle$ and $\langle J_{23} \rangle$ with positive $\langle J_{13} \rangle$ gives rise to a counterclockwise (CCW) internal circulation if viewed from above. At $k=k_{3}$, the local thermal currents in all the links are the same and flow from hot to cold. When all local thermal currents flow in the same direction, we call it a unidirectional (UD) flow. When $k>k_3$, the local thermal currents on the $1-2$ and $2-3$ links flow from hot to cold, but the local thermal current on the $1-3$ link flows from cold to hot as indicated by the negative value. In this case, the local thermal currents give rise to a clockwise (CW) circulation. For $k>>k_{3}$, the local thermal currents become unidirectional again. The three patterns (CCW, UD, and CW) are illustrated in the top panel of Fig.~\ref{fig:redfield_phase}.
We confirm that although a local thermal current may flow from cold to hot in the steady state, the total steady-state thermal current is always from hot to cold, consistent with the second law of thermodynamics.
We emphasize that all the results are the steady-state values according to the RQME, not transient behavior. Together with the demonstration in classical systems~\cite{dugar2019geometry}, the geometry-based circulation of thermal current emerges in both quantum and classical regimes.

Fig. \ref{fig:redfield_phase} shows the phase diagram of local-flow patterns on the $k/k_{3}$ and $\epsilon/\omega_0$ plane. There are three regimes exhibiting the CCW circulation, unidirectional flow, and CW circulation, respectively. If  $\epsilon/\omega_0$ is small, both types of circulation are observable. The circulation has the property that the atypical local current is along the link with the smaller value of the harmonic coupling constant. For example, $k/k_3 >1$ implies the atypical current is along the link with $k_3$, which is the $1-3$ link, giving rise to CW circulation.  However, as $\epsilon/\omega_0$ is increased, the regimes of both circulations shrink. Beyond a threshold value of $\epsilon/\omega_0$, only the UD flow can survive.

\begin{figure}[t]
	\centering
	\includegraphics[width=0.48\textwidth]{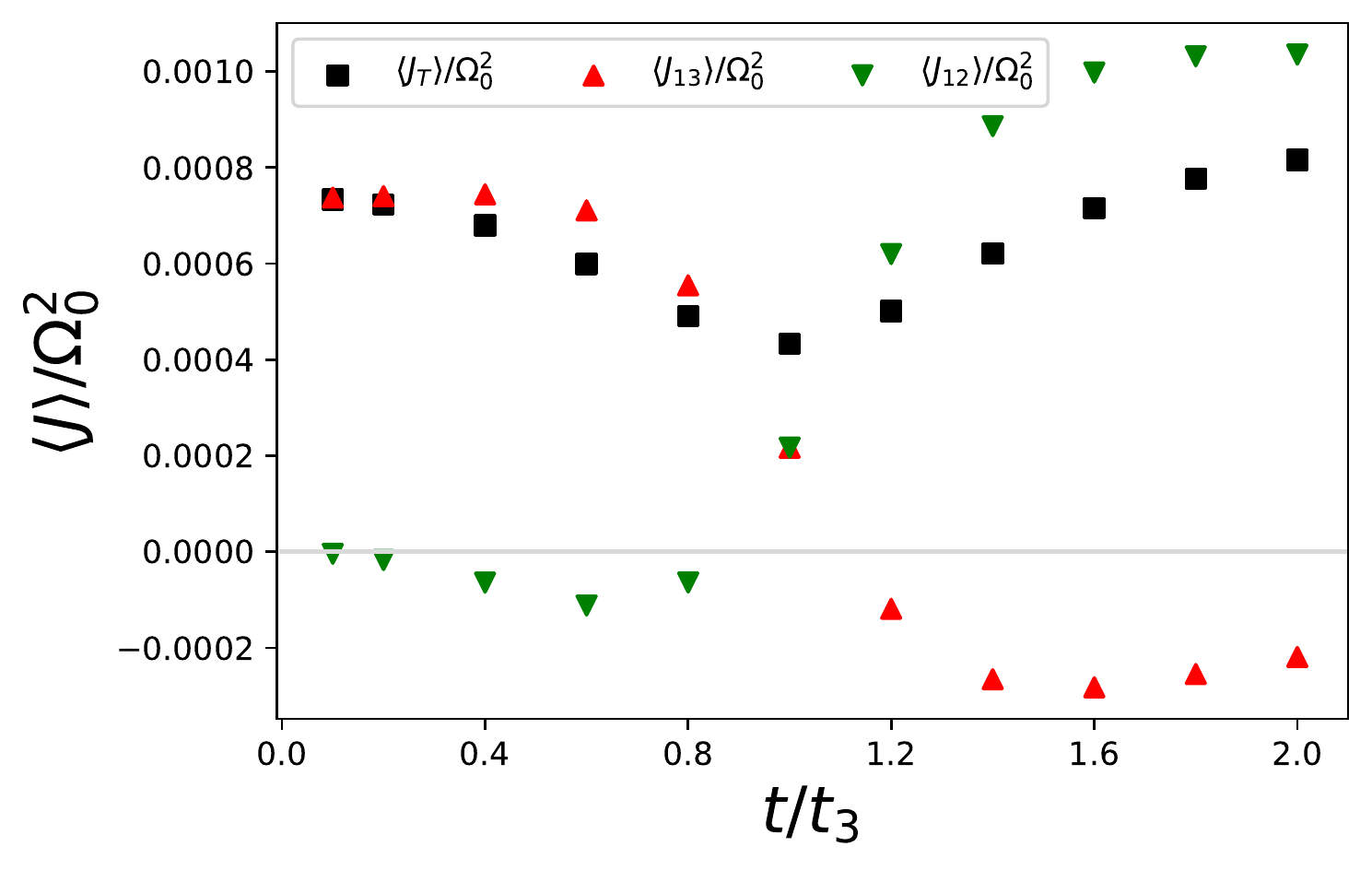}
	\caption{\label{fig:nonintlin3rdQlocalandtot} Total and local steady-state thermal currents of the non-interacting BHM described by the LQME as a function of $t/t_3$ for $t_3/\Omega_0=0.1$ and $\gamma/\Omega_0=0.1$ with $T_L/\Omega_0=1.01, T_R/\Omega_0=0.99$. $\Omega_0$ is the onsite frequency. The solid grey line marks the zero of the $y$-axis.}
\end{figure}

\subsection{Bose Hubbard model with LQME}
Here we present the local steady-state thermal currents from the LQME of the BHM of phonons illustrated in Fig. \ref{fig:schematic_red_lin} (b). We begin with the non-interacting BHM of Eq. \eqref{eq:BHHam} with $U=0$. Within the LQME \eqref{eq:LQME}, the local thermal currents can be evaluated once the reduced density matrix is obtained. For the noninteracting BHM, we employ the third quantization formalism to obtain the steady-state correlations from the LQME and then extract the information of the thermal currents associated with heat in thermodynamics.   
To stay in the regime where the rotating wave approximation applies, We take $t_3/\Omega_0=0.1$ and vary $t/\Omega_0$ from $0.01$ to $0.2$ in the following discussion. The system is coupled to two thermal reservoirs maintained at $T_L/\Omega_0=1.01$ and $T_R/\Omega_0=0.99$, with symmetric system-reservoir coupling constants $\gamma_L=\gamma_R=\gamma$.

Fig. \ref{fig:nonintlin3rdQlocalandtot} shows the total and local currents as a function of $t/t_3$ for $\gamma/\Omega_0=0.1$. As we vary $t/t_3$, the local currents follow a similar trend observed in the case of the harmonically coupled quantum oscillators described by the RQME shown in Fig.~\ref{fig:redlocalandtot}. The negative values of the local thermal currents indicate the emergence of a geometry-based circulation. For $t/t_3<1$ ($t/t_3>1$), a CCW (CW) circulation is present in the steady state. Around $t/t_3=1$, all the local currents flow in the same direction. Although the total thermal current remains positive and is consistent with the second law of thermodynamics, the local circulation would have been overlooked if only the total thermal current is reported. We also notice that the total current exhibits a dip when the circulation patterns changes from CW to CCW or vice versa, a phenomenon already found in thermal transport of classical harmonic systems~\cite{dugar2019geometry}.

\begin{figure}[t]
	\centering
	\includegraphics[width=0.49\textwidth]{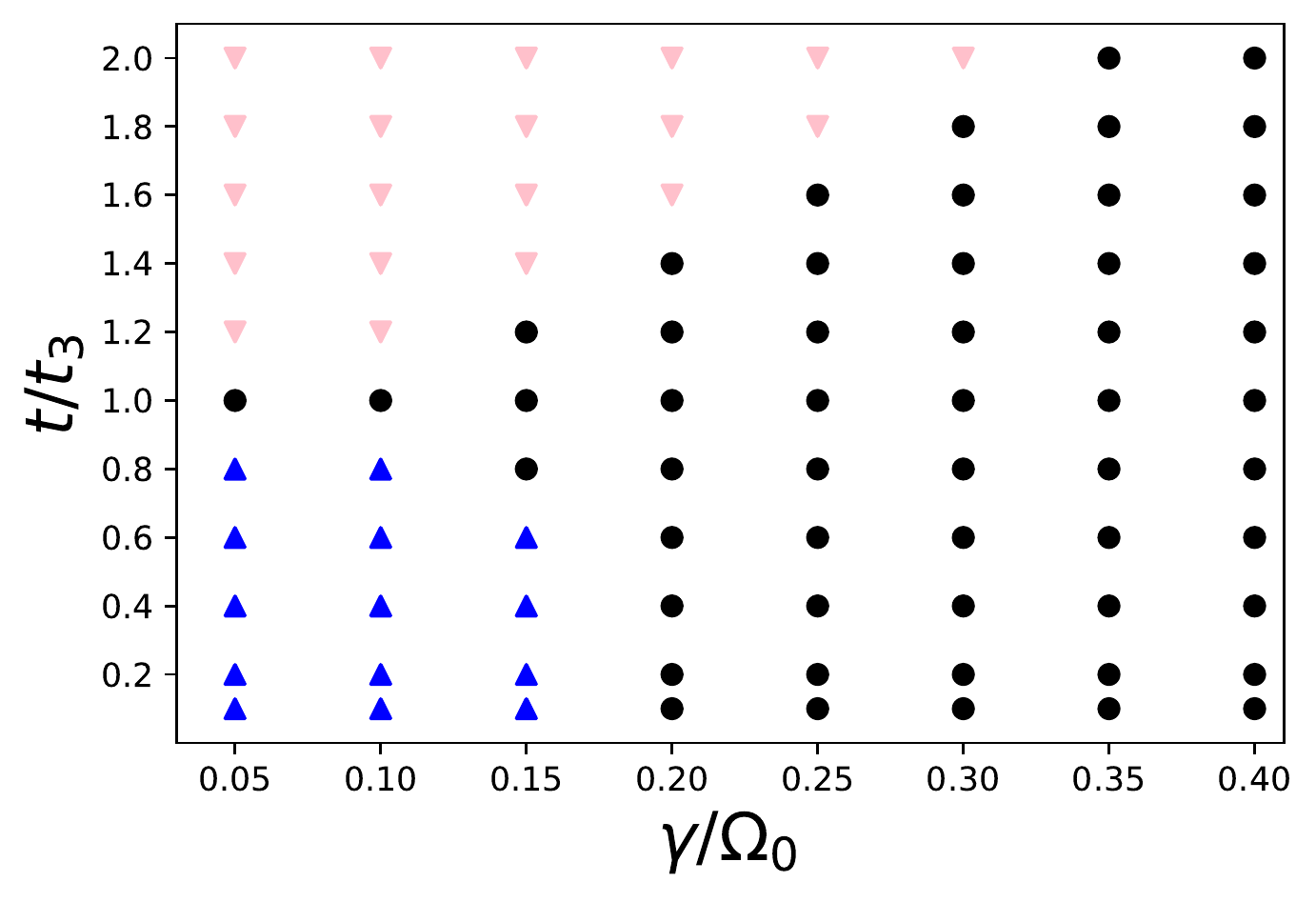}
	\caption{\label{fig:linphasemap} Phase diagram of the noninteracting BHM described by the LQME, showing the CCW circulation (blue triangles), CW circulation (pink inverted triangles), and UD flow (black dots) on the plane of $t/t_3$ and $\gamma/\Omega_0$. Here $T_L/\Omega_0=1.01, T_R/\Omega_0=0.99$, and $t_3/\Omega_0=0.1$.}
\end{figure}

After identifying the three patterns (CW, CCW, and UD) of the local thermal currents, we present the phase diagram of the noninteracting BHM described by the LQME in a multi-path geometry as a function of $t/t_3$ and $\gamma/\Omega_0$ in Fig. \ref{fig:linphasemap}. Both the CCW and CW circulations survive in the regime with small $\gamma/\Omega_0$. As $\gamma/\Omega_0$ increase, the regions of both circulations decrease, similar to the RQME results of Sec. \ref{res:RQME}. We note that the CCW circulation regime disappears more rapidly compared to the CW circulation regime as $\gamma/\Omega_0$ increases. 

So far, the geometry-based circulation has been demonstrated in quadratic or noninteracting quantum systems. In the following, we use the BHM in the LQME as a concrete example to show that the geometric mechanism is robust against nonlinear interactions. In the weakly interacting regime, $U/\Omega_0$ is the smallest energy scale, satisfying $U < min(t,t_3)$, where $min(t,t_3)$ denotes the smaller one of $t$ and $t_3$. In the presence of nonlinear interactions, the thermal steady-state current depends on the four-operator correlations in addition to the usual two-operator correlations, as shown in Eq. \eqref{eq:localcurrentwithU}. The four-operator correlations implicitly depend on the number density of the phonons, making it a nonlinear problem that is sensitive to the system configuration. 

\begin{figure}[t]
	\centering
	\includegraphics[width=0.48\textwidth]{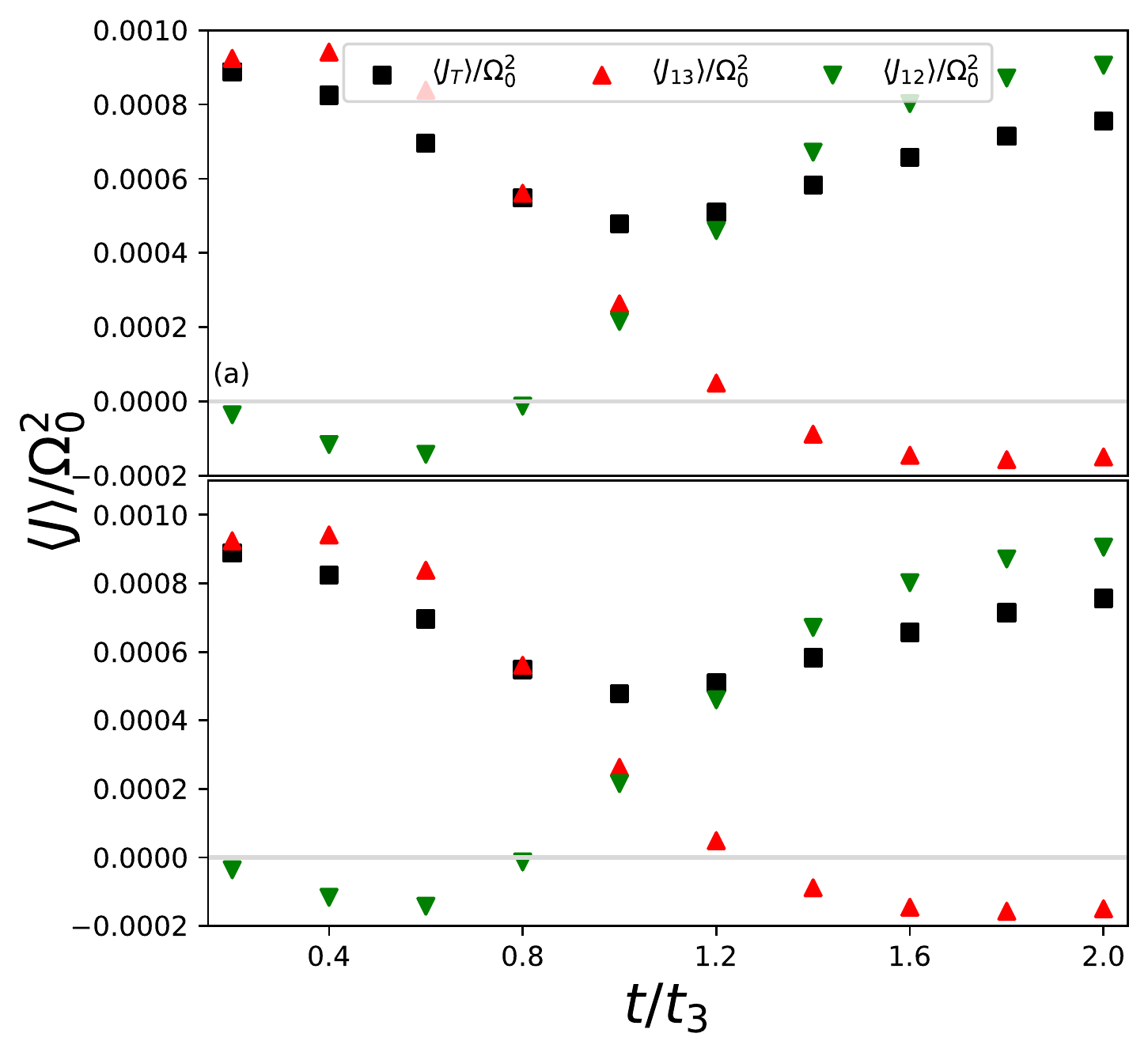}
	\caption{\label{fig:intlinlocalandtot} Total and local steady-state thermal currents of the interacting BHM described by the LQME for (a) $U/\Omega_0=0.01$ and (b) $U/\Omega_0=0.05$ as a function of $t/t_3$ with standardized tunneling coefficients satisfying $min(t,t_3)/\Omega_0=0.1$. Here $T_L/\Omega_0=1.01, T_R/\Omega_0=0.99$, and $\gamma/\Omega_0=0.1$. The grey horizontal line marks the zero of the vertical axis.
	}
\end{figure}

The local and total steady-state thermal currents of the BHM described by the LQME with interaction strength $U/\Omega_0=0.01$ and $0.05$ are shown in Fig. \ref{fig:intlinlocalandtot}. The maximal phonon number per site is limited to $4$. We notice that the transient time before the system reaches the steady state increases substantially with $U/\Omega_0$, so we stay in the weakly interacting regime. To ensure $U/\Omega_0$ remains the smallest energy scale in the problem, we use the standardized tunneling coefficients in Fig. \ref{fig:intlinlocalandtot}. Explicitly, we choose the tunneling coefficients so that $min(t,t_3)/\Omega_0=0.1$ to ensure $min(t,t_3)>U$. We have verified that the net current of the site not connected to the reservoirs is zero, so there is no accumulation of energy in the system in the steady state. Moreover, the signs of the local currents $\langle J_{12} \rangle$ and $\langle J_{23} \rangle$ agree, allowing a consistent identification of the patterns of local currents. 

Importantly, Fig. \ref{fig:intlinlocalandtot} shows that both circulation patterns survive in the interacting BHM with a multi-path geometry. From Figs. \ref{fig:linphasemap} and \ref{fig:intlinlocalandtot}, one can see that the weakly-interacting systems behave qualitatively the same as the noninteracting system. However, Fig. \ref{fig:linphasemap} is from the noninteracting system with the full Fock space while Fig. \ref{fig:intlinlocalandtot} is from the interacting BHM with a cap on the maximal phonon number per site (set to $4$ in Fig.~\ref{fig:intlinlocalandtot}). By solving the LQME of an effective BHM of photons, Ref.~\cite{dugar2020geometry} shows quantitative dependence of the geometry-based circulation regimes on the maximal particle number per site. Here we have checked that the variation of the circulation regimes with the maximal particle number is gradual without qualitative changes in phononic transport as well.
The BHM in the LQME thus offers an explicit example confirming that the geometry-based circulation is not unique to noninteracting or quadratic systems. Moreover, we have verified that the geometry-based circulation survives in the presence of asymmetric system-reservoir couplings ($\gamma_L\neq \gamma_R$). We mention that a suppression of the geometry-based circulation by nonlinear interactions has been demonstrated in electronic \cite{lai2018tunable} and photonic \cite{dugar2020geometry} transport. Importantly, the demonstrations of the geometry-based circulation in fermionic systems~\cite{lai2018tunable} and bosonic systems presented here and in Ref. \cite{dugar2020geometry} show the mechanism also transcends the spin statistics.

\section{Experimental implications and possible applications}\label{sec:exp}
The systems shown in Fig. \ref{fig:schematic_red_lin} for studying geometric effects in quantum thermal transport may be realizable in quantum dots \cite{noiri2017triangular}, optomechanical systems \cite{luk2011anomalous, yang2020phonon}, and atomic or molecular systems \cite{markussen2013phonon, evangeli2013engineering, thomas2017quantum} with suitable arrangements in a multi-path geometry and proper thermal reservoirs. In accordance with the theoretical parameters, the geometry-based circulations may survive in experiments performed at liquid helium temperature with the frequencies of the system of interest in the terahertz range. 
The setups may need the reservoirs to be connected to a specified part of the system, which may be achieved by focused laser 
that pumps or dissipates energy in a selected region.
There have been tremendous progresses in manipulating atoms with light. For example, Ref. \cite{yang2020phonon} studies phonon transport in two SiN nanomechanical resonators coupled to a cavity field, Refs. \cite{thomas2017quantum, barter2018quantum} trap ultracold atoms at the interference of coherent light beam, Refs. \cite{liu2018building,liu2019molecular} assemble single molecules from atoms through optical tweezers, Ref. \cite{beugnon2007two} demonstrates coherent transport of a neutral atom, and many others. Those techniques may be allow further studies of geometric effects in quantum transport.
For nanomechanical or molecular systems, the thermal currents may be measured through thermoreflectance  \cite{el2021heat} or scanning-probe techniques \cite{cui2017perspective, mosso2019thermal, evangeli2013engineering, meier2014length}. The systems shown in Fig. \ref{fig:schematic_red_lin} may be modified to function as a quantum thermal transistor by adding a reservoir to the second oscillator, as proposed in Refs. \cite{joulain2016quantum,  guo2019multifunctional, wang2018heat}.

There has been a proposal of thermal memory elements by recording information via temperature, but those devices usually require a non-linear element \cite{wang2008thermal, li2012colloquium}. For example, the bistable states of a non-linear one-dimensional chain made of two Frenkel-Kontorova segments may be utilized as a phononic memory \cite{wang2008thermal}. Here, we propose a different type of thermal memory element, which stores data by recording the types of local-flow patterns via the geometric effect. The geometry-based thermal memory does not necessarily requires non-linearity or additional manipulation of local temperature. In the geometry-based design, any two flow patterns (CCW, UD, and CW) can act as a bistable state for recording the binary digits 0 and 1. As shown previously, the flow patterns can be tuned by a variety of internal and external parameters. The bistability of the geometry-based system may also find applications in thermal switching or thermal routing in quantum systems because the magnitude and direction of the local thermal currents are highly controllable. A crucial step towards the realization of those applications is the control of the internal and external parameters, which may be achieved  electromagnetically \cite{kurt2020first} via lasers or mechanically \cite{torres2015tuning, majumdar2015vibrational} via stress or strain. Reading out the information or status of the system is another challenge and may be performed by introducing additional or adjacent thermoelectric elements to siphon out some phonons and convert the information to electric signals for performing the measurement \cite{li2012colloquium, dubi2011colloquium, ye2016thermodynamic, rojo2013review} 

\section{\label{sec:level4}Conclusion}
By analyzing concrete examples of transport in harmonically coupled quantum oscillators described by the RQME and the BHM of phonons described by the LQME with multi-path geometries, we have demonstrated the geometry-based circulation of local thermal currents in the steady state with a local thermal current flowing against the total current. The geometry-based mechanism is insensitive to the details of the Hamiltonian and the modeling of the reservoir and system-reservoir coupling. The ubiquity of the geometry-based circulation of thermal currents in quantum and classical systems shows that the phenomenon transcends the classical and quantum boundary. Nevertheless, the total thermal current always flows from hot to cold, respecting the second law of thermodynamics and failing to reflect the local atypical behavior. Importantly, the patterns of the local thermal current are tunable by internal parameters of the system and external system-reservoir or system-substrate couplings. Moreover, the geometry-based circulation of the local steady-state thermal current is robust against nonlinear interactions. Geometric effects thus offer more alternatives to manipulate thermal transport in quantum systems.

\acknowledgements{This work was supported by the NSF under Grant No. PHY-2011360. The simulations were performed on the MERCED cluster funded by NSF under Grant No. ACI-1429783.}

\appendix
\section{Third quantization formalism for the RQME}\label{appendix}
The third quantization method formulates a Fock space for the quantum operators by using left- and right- multiplication maps in the operator space. We follow Ref. \cite{vzunkovivc2012heat}, which applies the formalism to solve the RQME and derives the Lyapunov equation for the nonequilibrium steady-state (NESS) coordinate-momentum correlation matrix. 
To begin, we diagonalize the matrix $\boldsymbol{Q}$ in Eq. \eqref{eq:QOham_in_Q}:
\begin{equation}
    \boldsymbol{Q}=\boldsymbol{U}\boldsymbol{\Omega}\boldsymbol{U^{\dagger}}.
\end{equation}
Here the columns of $\boldsymbol{U}$ are the right eigenvectors of $\boldsymbol{Q}$ and $\boldsymbol{\Omega}$ is a diagonal matrix whose entries are the corresponding eigenvalues of $\boldsymbol{Q}$. We also define $\lambda_j=\sqrt{\boldsymbol{\Omega}_{j,j}}$. For the quantum-oscillator system, we can transform the momentum and coordinate vectors to the normal basis:
\begin{eqnarray}
\underline{p'}=\underline{p}.\boldsymbol{U}  \qquad \underline{q'}=\underline{q}.\boldsymbol{U}
\end{eqnarray}

Hereon, the prime $'$ indicates the vectors and operators in the normal basis. According to Ref. \cite{vzunkovivc2012heat}, we need to solve the following continuous Lyapunov equation:
\begin{equation}\label{eq:LyapunovEqn}
    \boldsymbol{X}^T\boldsymbol{Z}+\boldsymbol{Z}\boldsymbol{X}=\boldsymbol{Y}.
\end{equation}
The solution yields the matrix $\boldsymbol{Z}$, whose elements are equivalent to the NESS correlation functions $ {\boldsymbol{Z}_{ij}^{a,b}}=\langle a_ib_j \rangle$, with $a,b \in \{p,q\}$. $X^T$ and $Y$ in Eq. \eqref{eq:LyapunovEqn} are defined as
\begin{equation}
\boldsymbol{X}^T= \left(\begin{array}{cc} 
\boldsymbol{M_{im}^{q'p'}} & \frac{\boldsymbol{\Omega}}{2} \\
-\frac{\boldsymbol{\mathbb{1}_3}}{2} & \boldsymbol{0}
\end{array}
\right)
\end{equation}
and 
\begin{equation}
\boldsymbol{Y}= \frac{1}{2}\left(\begin{array}{cc} 
\boldsymbol{M_{r}^{q'q'}}+(\boldsymbol{M_{r}^{q'q'}})^T & \boldsymbol{0} \\
\boldsymbol{0} & \boldsymbol{0}
\end{array}
\right).
\end{equation}
Here $M^{\alpha\beta}$ are the matrices encoding the influence of the reservoirs, whose subscripts $im$ and $r$ refer to the imaginary and real parts, respectively. Following, Ref. \cite{vzunkovivc2012heat}, the matrices are
\begin{widetext}
\begin{equation}
    \boldsymbol{M^{q'q'}}= \frac{1}{2}X'_{L}\otimes X'_{L}diag((1+\exp(\frac{\lambda_{j}}{T_L}))\Gamma_{L}(\lambda_j)) + \frac{1}{2}X'_{R}\otimes X'_{R}diag((1+\exp(\frac{\lambda_{j}}{T_R}))\Gamma_{R}(\lambda_j)),
\end{equation}
\begin{equation}
    \boldsymbol{M^{q'p'}}= \frac{\iota}{2}X'_{L}\otimes X'_{L}diag((\frac{\exp(\lambda_{j}/T_L)-1}{\lambda_j})\Gamma_{L}(\lambda_j)) + \frac{\iota}{2}X'_{R}\otimes X'_{R}diag((\frac{\exp(\lambda_{j}/T_R)-1}{\lambda_j})\Gamma_{R}(\lambda_j)).
\end{equation}
\end{widetext}
The steady-state local thermal currents and occupation numbers can then be calculated from the NESS correlation functions.


\section{Some LQMEs from RQME}\label{app:QME}
In the secular approximation (equivalent to the rotating wave approximation in this case), the interaction-picture equivalent of Eq.~\eqref{eq:RQME} is written in the energy eigenbasis of the system Hamiltonian. The difference of the eigen-frequencies $(\omega^{'}-\omega^{''})$ are called the Bohr frequencies. The RQME has all the time-dependent terms in the form of $ e^{i(\omega^{'}-\omega^{''})t}$. The secular approximation averages over the fast oscillating terms. For the RQME in the energy eigenbasis of the system Hamiltonian, the approximation essentially removes the terms whose associated frequencies are different $(\omega^{'} \neq \omega^{''})$ \cite{vogt2013stochastic, lidar2019lecture}, resulting in a global LQME. However, the approximation is not satisfied if some energy eigenvalues or the Bohr frequencies are nearly degenerate, so the global LQME only works for systems with well separated eigenvalues, $(\omega^{'} - \omega^{''})> \gamma$, where $\gamma$ is the system-reservoir coupling constant.  

On the other hand, performing the weak internal coupling approximation \cite{wichterich2007modeling, purkayastha2016out} on the RQME leads to a local LQME. Let $t_{ij}$ denote the tunneling coefficient between oscillators $i$ and $j$. Ref. \cite{purkayastha2016out} imposes the equivalent of $\gamma > t_{ij}$ and uses a perturbative expansion with respect to $t_{ij}$ to obtain a local LQME. Refs. \cite{wichterich2007modeling, asadian2013heat} require the equivalent of $\Omega_0 > t_{ij}$ such that the time dependence of $X_{L,R}(-\tau)$ in Eq. \eqref{eq:redfielddiss} can be approximated by the "local Hamiltonian" from the $\Omega_0$-term instead of the full non-interacting BHM.



\bibliographystyle{apsrev}

\end{document}